\documentclass[prd,showpacs]{revtex4}
\usepackage{amsmath,amssymb}

\numberwithin{equation}{section}
\begin{document}
\title{Thermal effects on the Casimir energy of a Lorentz-violating scalar in magnetic field}
\author{Andrea Erdas}
\email{aerdas@loyola.edu}
\affiliation{Department of Physics, Loyola University Maryland, 4501 North Charles Street,
Baltimore, Maryland 21210, USA}
\begin {abstract} 
In this work I investigate the finite temperature Casimir effect due to a massive and charged scalar field that breaks Lorentz invariance in a CPT-even, aether-like way. I study the cases of Dirichlet and mixed (Dirichlet-Neumann) boundary conditions on a pair of parallel plates. I will not examine the case of Neumann boundary conditions since it produces the same results as Dirichlet boundary conditions. The main tool used in this investigation is the $\zeta$-function technique that allows me to obtain the Helmoltz free energy and Casimir pressure in the presence of a uniform magnetic field perpendicular to the plates. Three cases of Lorentz asymmetry are studied: timelike, spacelike and perpendicular to the magnetic field, spacelike and parallel to the magnetic field. Asymptotic cases of small plate distance, high temperature, strong magnetic field, and large mass will be considered for each of the three types of Lorentz asymmetry and each of the two types of boundary conditions examined. In all these cases simple and very accurate analytic expressions of the thermal corrections to the Casimir energy and pressure are obtained and I discover that these corrections strongly depend on the direction of the unit vector that produces the breaking of the Lorentz symmetry.
\end {abstract}
\maketitle
\section{Introduction}
\label{1}
The first to predict an attractive force between two uncharged parallel plates due to quantum effects, was Hendrik Casimir \cite{Casimir:1948dh}. Sparnaay's experiments \cite{Sparnaay:1958wg} ten years later tested this phenomenon, quite difficult to observe, and were relatively consistent with it. Many increasingly accurate and sophisticated experiments followed throughout the decades \cite{Bordag:2001qi,Bordag:2009zz}, fully confirming the Casimir effect, caused by the quantum fluctuations of the electromagnetic field in vacuum. Quantum fluctuations of other vector, scalar and fermion fields also cause Casimir forces. These forces are strongly dependent on the boundary conditions at the plates of the quantum field involved. In the case of a scalar field, Dirichlet or Neumann boundary conditions cause an attractive force between the two parallel plates, while mixed (Dirichlet-Neumann) boundary conditions cause a repulsive force \cite{Boyer:1974}. Casimir forces display also a strong dependence on the geometry of the plates. The repulsive nature of the spherical Casimir effect \cite{Boyer:1968uf} is caused by the electromagnetic field, or by a scalar field, satisfying Dirichlet boundary conditions at the spherical shell.

In recent years the violation of Lorentz symmetry in quantum field theories has been investigated with the proposal of models where such a violation causes a space-time anisotropy \cite{Ferrari:2010dj,Ulion:2015kjx}. Several Lorentz symmetry breaking mechanism have been presented within the context of quantum gravity \cite{Alfaro:1999wd,Alfaro:2001rb}, of string theory where a spontaneous violation of Lorentz invariance at the Planck scale has been conjectured \cite{Kostelecky:1988zi}, and of quantum field theories with varying coupling constants \cite{Anchordoqui:2003ij, Bertolami:1997iy, Kostelecky:2002ca}. The repercussions of a space-time anisotropy on the Casimir effect have been analyzed for the case of a real scalar field in vacuum \cite{Cruz:2017kfo} and at finite temperature \cite{Cruz:2018bqt}, where the scalar field is in equilibrium with a thermal bath. These two papers \cite{Cruz:2017kfo,Cruz:2018bqt} investigate the Casimir effect due to a modified Klein-Gordon field that breaks the Lorentz symmetry in a CPT-even, aether-like manner. The consequence of space-time anisotropy have also been studied extensively in the case of Lorentz-breaking extensions of QED \cite{Frank:2006ww, Kharlanov:2009pv, Martin-Ruiz:2016lyy}. The Casimir effect is one of the most studied quantum phenomena, theoretically and experimentally, and has been proposed by all these authors as a laboratory for studying Lorentz symmetry violation.

Authors have investigated magnetic corrections to the Casimir effect caused by a charged scalar field in the standard Lorentz symmetric space-time \cite{CougoPinto:1998td}, and also the combination of thermal and magnetic effects within standard space-time \cite{CougoPinto:1998jg, Erdas:2013jga, Erdas:2013dha}, and recently a paper has been published that investigates vacuum magnetic corrections to the Casimir effect of a scalar field that breaks Lorentz symmetry \cite{Erdas:2020ilo} however, there has not been a study of the combination of thermal and magnetic corrections to the Casimir effect of a scalar field that breaks Lorentz invariance. This paper, a sequel to my previous work \cite{Erdas:2020ilo}, intends to fill that gap and study the effect of a constant magnetic field $\bf B$, perpendicular to the plates, on the Casimir energy and pressure caused by a scalar field that breaks Lorentz symmetry and is in thermal equilibrium with a heat reservoir at finite temperature $T$. I will study a charged scalar field that breaks the Lorentz symmetry in the manner proposed by Refs. \cite{Cruz:2017kfo,Cruz:2018bqt} and satisfies either Dirichlet or mixed boundary conditions at two square parallel plates a distance $a$ from each other.

When two parallel plates face each other in vacuum and a quantum field in its ground state, vanishing at the plates, permeates the space, an infinite number of wavelengths contributes to the calculation of the vacuum energy producing an infinite value. In order to obtain a finite quantity, one must subtract the vacuum energy of that space without the plates, from the vacuum energy of that same space with the plates present. Both quantities are infinite but, through a careful regularization procedure, one is able to obtain a finite quantity, the Casimir energy, which does not depend on the regularization method. In this work I will use a regularization procedure introduced by Hawking \cite{Hawking:1976ja}, the zeta function technique, extensively used in the computation of vacuum Casimir energies \cite{Elizalde:1988rh, Elizalde:2007du} and also in the context of finite temperature Casimir energy computations \cite{Santos:1998vb, CougoPinto:1998xn, Santos:1999yj}. Other regularization methods are available and widely used in the literature when computing Casimir energies, such as the Abel-Plana method used in Refs. \cite{Cruz:2017kfo,Cruz:2018bqt}.

In Sec. \ref{2} of this paper I illustrate the model of a scalar field that breaks Lorentz symmetry in an aether-like and CPT-even way. In Sec. \ref{3}, I impose Dirichlet boundary conditions at the plates for the scalar field and, in subsection A, examine the case of timelike anisotropy. I derive the zeta function and, from it, obtain the Casimir energy in the asymptotic limits of small plate distance, high temperature, strong magnetic field and large mass. In subsections B and C I do the same for the cases of spacelike anisotropy in the plane perpendicular to $\bf B$ and spacelike anisotropy in the direction of $\bf B$, respectively. In Sec. \ref{4} I investigate mixed (Dirichlet-Neumann) boundary conditions at the plates and obtain the zeta function and the asymptotic limits of the Casimir energy listed above, for each of the three cases of space-time anisotropy. In Sec. \ref{5} I obtain and discuss the Casimir pressure for all these scenarios. My conclusions and a discussion of the results obtained in this paper are presented in Sec. \ref{6}. In Appendix I, I show the details of the calculation of the high temperature limit of the zeta function for Dirichlet boundary conditions and timelike asymmetry. In Appendix II, I show the calculation of the high temperature limit of the zeta function for mixed boundary conditions and timelike asymmetry.

\section{The model}
\label{2}

In this paper I study the Casimir effect caused by a scalar field $\phi$ of charge $e$ and mass $M$ that violates Lorentz symmetry. The theoretical model of a scalar field that breaks Lorentz symmetry in an aether-like and CPT-even manner was introduced in Ref. \cite{Cruz:2017kfo}. In such a model, the derivative of the scalar field is coupled to a fixed unit four-vector $u^\mu$, thus causing the Lorentz symmetry violation. This complex scalar field obeys the following Klein Gordon equation
\begin{equation}
[\Box 
+\lambda(u\cdot\partial)^2+M^2]\phi=0,
\label{KG}
\end{equation}
where the unit four-vector $u^\mu$ points in the direction in which the Lorentz symmetry is broken and the dimensionless parameter $\lambda \ll 1$ determines the amount of symmetry violation. I will investigate two square parallel plates of side $L$, perpendicular to the $z$ axis and located at $z=0$ and $z=a$, in thermal equilibrium with a heat reservoir at temperature $T$ and will obtain thermal corrections to their Casimir energy  caused by the aforementioned Lorentz violating scalar field. A uniform magnetic field $\bf B$ pointing in the $z$ direction will also be included in this study. For such a system the imaginary time formalism of finite temperature field theory is convenient, and it allows only field configurations satisfying the following periodic condition
\begin{equation}
\phi(x, y, z,\tau)=\phi(x, y, z,\tau+\beta),
\label{temperature}
\end{equation}
for any $\tau$, where $\beta=1/T$ is the periodic length in the Euclidean time axis. In addition to the finite temperature boundary condition
(\ref{temperature}), I will impose either Dirichlet or mixed boundary conditions at the plates and will use the generalized zeta function technique to study this system. Neumann boundary conditions will not be investigated, since they produce the same result as Dirichlet. I will investigate a unit four-vector $u^\mu$ that is timelike, spacelike and perpendicular to $\bf B$, and spacelike and parallel to $\bf B$.

The Helmholtz free energy of the scalar field is
$$F=\beta^{-1}\log \,\det \left(D_{\rm E}|{\cal F}_a\right),$$
where $D_{\rm E}$ is the Klein Gordon operator in Euclidean time and the symbol ${\cal F}_a$ indicates the set of eigenfunctions of $D_{\rm E}$ which satisfy condition 
(\ref{temperature}),  and either Dirichlet or mixed boundary conditions at the plates. In the absence of a magnetic field, $D_{\rm E}$ is given by
\begin{equation}
D_E=-{\frac{\partial^2}{ \partial \tau^2}}-\nabla^2 
+\lambda(u\cdot\partial)^2+M^2.
\label{KGoperator}
\end{equation}
Notice that at this initial stage the magnetic field is not present and it will be introduced later.
When $u^\mu$ is timelike and the finite temperature condition of Eq. (\ref{temperature}) is imposed, the eigenvalues of $D_E$ are:
\begin{equation}
\left\{(1+\lambda)\frac{4\pi^2 m^2}{ \beta^2}+k_x^2+k_y^2+k_z^2+M^2\right\},
\label{eigenvalues1}
\end{equation}
where $k_x, k_y \in \Re$, $m=0,\pm 1,\pm 2,\cdots$, and $k_z$ will take different discrete values if $\phi$ satisfies Dirichlet or mixed boundary conditions at the plates.
When $u^\mu$ is spacelike and lies in the $x$-$y$ plane, I take $u^\mu= \left(0, \frac{1}{ \sqrt{2}} ,  \frac{1}{ \sqrt{2}}  ,0\right)$ and the eigenvalues of $D_E$ are:
\begin{equation}
\left\{\frac{4\pi^2 m^2}{ \beta^2}+\left(1-\frac{\lambda}{2}\right)(k_x^2+k_y^2)+k_z^2+M^2\right\}.
\label{eigenvalues2}
\end{equation}   
Finally, if $u^\mu$ is in the $z$ direction, I find the following eigenvalues of $D_E$:
\begin{equation}
\left\{\frac{4\pi^2 m^2}{ \beta^2}+k_x^2+k_y^2+\left(1-{\lambda}\right)k_z^2+M^2\right\}.
\label{eigenvalues3}
\end{equation}

Introducing the magnetic field modifies the eigenvalues of Eqs. (\ref{eigenvalues1}, \ref{eigenvalues2}, \ref{eigenvalues3}) by changing $k^2_x+k^2_y$ into $2eB(\ell+\frac{1}{ 2})$, where $\ell=0,1,2,\cdots$, labels the Landau levels. 

In the next sections I will construct the zeta function, evaluate it and obtain the Casimir energy and its thermal corrections for Dirichlet and mixed boundary conditions in the three cases of timelike Lorentz anisotropy, spacelike anisotropy in the $x$-$y$ plane, and spacelike anisotropy in the $z$ direction.
\section{Dirichlet boundary conditions}
\label{3}
Dirichlet boundary conditions constrain the field $\phi$ to vanish at the plates
\begin{equation}
\phi(x, y, 0,\tau)=\phi(x, y,a,\tau) = 0,
\label{Dirichlet}
\end{equation}
so in this case
\begin{equation}
k_z=\frac{n\pi}{ a},
\label{Dirichlet2}
\end{equation}
with $n=1,2,3,\cdots$. In the rest of this section I will investigate a Lorentz violating scalar field satisfying Dirichlet boundary conditions at the plates in the presence of a uniform magnetic field ${\bf B} = B {\hat z}$ and, in subsection \ref{3_2}, will calculate the vacuum and thermal part of its free energy (i.e. Casimir energy) for the case of timelike Lorentz anisotropy. The cases of spacelike anisotropy in the $x$-$y$ plane and spacelike anisotropy in the $z$ direction will be examined in subsections \ref{3_3} and \ref{3_4}, respectively.
\subsection{Timelike anisotropy}
\label{3_2}
I will now construct the zeta function for the case of timelike Lorentz anisotropy. Using the eigenvalues (\ref{eigenvalues1}), replacing $k^2_x+k^2_y$ with $2eB(\ell+\frac{1}{ 2})$ due to the presence of the magnetic field, and using Eq. (\ref{Dirichlet2}) for $k_z$,
I obtain
\begin{equation}
\zeta(s)=\mu^{2s}(1+\lambda)\sum_{m=-\infty}^\infty \sum_{n=1}^\infty \frac{L^2eB}{ 2\pi}\sum_{\ell=0}^\infty \left[(1+\lambda)\left(\frac{2\pi m}{ \beta}\right)^2+(2\ell+1)eB+\left(\frac{n\pi}{ a}\right)^2+M^2\right]^{-s},
\label{zeta1_1}
\end{equation}
where, as it is done routinely when using the zeta function technique, I use the parameter $\mu$ with dimension of mass \cite{Hawking:1976ja} 
 to keep $\zeta(s)$ dimensionless for all values of $s$. The factor $\frac{L^2eB}{2\pi}$ keeps track of the degeneracy of the Landau levels, and the extra factor of $(1+\lambda)$ is needed in the case of timelike asymmetry as shown in Ref. \cite{Cruz:2017kfo}, where it is shown that the vacuum energy, obtained by taking the vacuum expectation value of the Hamiltonian, carries this factor.

Using the following identities
\begin{equation}
z^{-s}=\frac{1}{ \Gamma(s)}\int_0^\infty dt\, t^{s-1}e^{-zt},
\label{z-s}
\end{equation}
\begin{equation}
\sum_{\ell=0}^\infty e^{-(2\ell +1)z}=\frac{1}{ 2 \sinh z},
\label{sinh}
\end{equation}
I write the zeta function as
\begin{equation}
\zeta(s)={\mu^{2s}(1+\lambda)\over\Gamma(s)}\frac{L^2eB}{ 4\pi}\int_0^\infty dt\, t^{s-1}{e^{-M^2t}\over\sinh eBt}\sum_{m=-\infty}^\infty  e^{-(1+\lambda){4\pi^2 m^2\over \beta^2}t}\sum_{n=1}^\infty e^{-{n^2\pi^2\over a^2}t},
\label{zeta1_2}
\end{equation}
where $\Gamma(s)$ is Euler's gamma function.
While it is not possible to evaluate this zeta function exactly, it is possible to obtain simple analytic expressions in several asymptotic regimes of the parameters.

I will first examine the asymptotic regime of small plate distance: $a^{-1}\gg T,M,\sqrt{eB}$. Applying the Poisson resummation method to the $m$-sum in Eq. (\ref{zeta1_2}), I find
\begin{equation}
\sum_{m=-\infty}^\infty  e^{-(1+\lambda){4\pi^2 m^2\over \beta^2}t} = {1\over 2T\sqrt{\pi(1+\lambda)t}}\sum_{m=-\infty}^\infty e^{-{m^2\over 4T^2(1+\lambda)t}},
\label{Poisson1}
\end{equation}
and can separate the zeta function into a vacuum part and a thermal part
\begin{equation}
\zeta(s)=\zeta_0(s)+\zeta_T(s).
\label{zeta1_3}
\end{equation}
The vacuum part of the zeta function is
\begin{equation}
\zeta_0(s)={\mu^{2s}\sqrt{1+\lambda}\over\Gamma(s)}{L^2\over 8\pi^{3\over 2}T}\int_0^\infty dt\, t^{s-{5\over 2}}{e^{-M^2t}}F(eBt)\sum_{n=1}^\infty e^{-{n^2\pi^2\over a^2}t},
\label{zeta0_1}
\end{equation}
and the thermal part is
\begin{equation}
\zeta_T(s)={\mu^{2s}\sqrt{1+\lambda}\over\Gamma(s)}{L^2\over 4\pi^\frac{3}{ 2}T}\int_0^\infty dt\, t^{s-{5\over 2}}{e^{-M^2t}}F(eBt)\sum_{n=1}^\infty e^{-{n^2\pi^2\over a^2}t} \sum_{m=1}^\infty e^{-{m^2\over 4T^2(1+\lambda)t}},
\label{zetaT_1}
\end{equation}
where I use the notation I introduced in Ref. \cite{Erdas:2020ilo}
\begin{equation}
F(z)=\frac{z}{ \sinh z}.
\label{F}
\end{equation}
The free energy $E$ is obtained easily from $\zeta(s)$
\begin{equation}
E=-T\zeta'(0),
\label{E}
\end{equation}
therefore, to obtain $E$, it is sufficient to evaluate $\zeta(s)$ when $s$ is near zero. I calculated the vacuum part of the zeta function in Ref. \cite{Erdas:2020ilo} and found that, in the asymptotic regime of small plate distance and for small $s$
\begin{equation}
\zeta_0(s)\simeq{\pi^2\sqrt{1+\lambda}L^2\over 8a^3T}\left[{1\over 90}-{M^2a^2\over 6\pi^2}+\left({M^4\over 2}-{e^2B^2\over 6}\right)\left({a\over \pi}\right)^4\left({1\over 2s}+\gamma_E  +\ln{\mu a\over 2\pi}\right)\right]s,
\label{zeta0_2}
\end{equation}
and, in the same paper, obtained the vacuum part of the Casimir energy
\begin{equation}
E_0=-{\pi^2\sqrt{1+\lambda}L^2\over 8a^3}\left[{1\over 90}-{M^2a^2\over 6\pi^2}+\left({M^4\over 2}-{e^2B^2\over 6}\right)\left({a\over \pi}\right)^4\left(\gamma_E  +\ln{\mu a\over 2\pi}\right)\right],
\label{E0_2}
\end{equation}
where $\gamma_E=0.5772$ is the Euler Mascheroni constant and the parameter $\mu$ takes the value $\mu = \sqrt{eB+M^2}$. 

The thermal part of the free energy is evaluated by making a change of integration variable from $t$ to ${ma\over 2\pi n T\sqrt{1+\lambda}}t$. In the asymptotic regime of small plate distance, $aT\ll1$, only the term with $n=m=1$ contributes significantly, and I obtain
\begin{equation}
\zeta_T(s)\simeq\left(\mu^2 a\over 2\pi T\sqrt{1+\lambda}\right)^{s}{(1+\lambda)^{\frac{5}{4}}L^2\over\Gamma(s)}{\sqrt{T\over 2a^{3}}}\int_0^\infty dt\, t^{s-{5\over 2}}e^{-{M^2a\over 2\pi T \sqrt{1+\lambda}}t}F\left({eBat\over 2\pi T \sqrt{1+\lambda}}\right) e^{-{\pi\over 2aT\sqrt{1+\lambda}}(t +1/t)}.
\label{zetaT_2}
\end{equation}
Using the saddle point method to evaluate the integral, I find
\begin{equation}
\zeta_T(s)=\left(\mu^2 a\over 2\pi T\sqrt{1+\lambda}\right)^{s}{(1+\lambda)^{\frac{3}{2}}L^2\over\Gamma(s)}{T\over a}e^{-{\pi\over a T \sqrt{1+\lambda}}}e^{-{M^2a\over 2\pi T \sqrt{1+\lambda}}}F\left({eBa\over 2\pi T \sqrt{1+\lambda}}\right),
\label{zetaT_3}
\end{equation}
and, using Eq. (\ref{E}), I obtain the thermal correction to the Casimir energy for small plate distance
\begin{equation}
E_T=-{(1+\lambda)^{\frac{3}{2}}L^2T^2\over a}e^{-{\pi\over a T \sqrt{1+\lambda}}}e^{-{M^2a\over 2\pi T \sqrt{1+\lambda}}}F\left({eBa\over 2\pi T \sqrt{1+\lambda}}\right).
\label{ET_1}
\end{equation}
In the case of small plate distance, $aT\ll 1$, the vacuum part of the Casimir energy is dominant when compared to the exponentially suppressed thermal correction.

Next I investigate the high temperature regime, $T\gg a^{-1}, \sqrt{eB}, M$. I apply the Poisson resummation method to the $n$-sum in Eq. (\ref{zeta1_2}) and obtain
\begin{equation}
\sum_{n=1}^\infty e^{-{n^2\pi^2\over a^2}t} = {a\over 2\sqrt{\pi t}}\sum_{n=-\infty}^\infty e^{-{n^2a^2\over t}}-{\frac{1}{2}},
\label{Poisson2}
\end{equation}
where the $-\frac{1}{2}$ term can be neglected since it is independent of $a$ and, when considered in the energy calculation, it produces the energy of a single plate \cite{Cruz:2018bqt} which is not relevant to this work. Poisson resumming the $n$-sum allows me to separate 
$\zeta(s)$ into a uniform energy density part, $\zeta_U(s)$, with a simple linear dependence on $a$
\begin{equation}
\zeta_U(s)={\mu^{2s}(1+\lambda)\over\Gamma(s)}{L^2a\over 8\pi^{3\over 2}}\int_0^\infty dt\, t^{s-{5\over 2}}{e^{-M^2t}F(eBt)}\left(1+2\sum_{m=1}^\infty  e^{-(1+\lambda){4\pi^2 m^2\over \beta^2}t}\right),
\label{zetaU_1}
\end{equation}
and a remaining part, not linear in $a$, that I indicate with $\zeta_{T, a}(s)$ and is
\begin{equation}
\zeta_{T,a}(s)={\mu^{2s}(1+\lambda)\over\Gamma(s)}{L^2a\over 4\pi^{3\over 2}}\int_0^\infty dt\, t^{s-{5\over 2}}{e^{-M^2t}F(eBt)}\left(1+2\sum_{m=1}^\infty  e^{-(1+\lambda){4\pi^2 m^2\over \beta^2}t}\right)\sum_{n=1}^\infty e^{-{n^2a^2\over t}}.
\label{zetaTa_1}
\end{equation}
In Appendix I, I show how to obtain $E_U$ from $\zeta_U(s)$ and find
\begin{eqnarray}
E_U=
&-&\!\!\!L^2a\left[{\pi^2(1+\lambda)^{5\over 2}\over 45}T^4-{(1+\lambda)^{3\over 2}\over 12}M^2T^2-{(1+\lambda)\over 2^{3\over 2}\pi}{(eB)^{3\over2}T}\zeta_H\left(-{\textstyle\frac{1}{2}} ; {\textstyle\frac{1}{2}}+{\textstyle\frac{M^2}{2eB}}\right)\right.
\nonumber \\
&+&\!\!\!
\left.
{(1+\lambda)^{1\over 2}\over 16\pi^2}\left({M^4}-{e^2B^2\over 3}\right)\left(\gamma_E-{\lambda\over 2}+\ln{\mu\over 4\pi T}\right)\right],
\label{EU_1}
\end{eqnarray}
where, again, $\mu = \sqrt{eB+M^2}$ and  $\zeta_H(s;x)$ is the Hurwitz zeta function defined in Appendix I. Clearly the dominant term here is the Stefan Boltzmann term, $-L^2a{\pi^2\over 45}(1+\lambda)^{5\over 2}T^4$, as expected. It is interesting to explore the zero mass limit and the zero magnetic field limit of $E_U$. We have 
$\zeta_H\left(-{\textstyle\frac{1}{2}} ; {\textstyle\frac{1}{2}}\right)=\left({\textstyle\frac{1}{\sqrt 2}}-1\right)\zeta_R\left(-{\textstyle\frac{1}{2}}\right)$, where the Riemann zeta function takes the value $\zeta_R\left(-{\textstyle\frac{1}{2}}\right)=-0.207886$, and therefore the zero mass limit of $E_U$ is
\begin{equation}
E_U=
-L^2a\left[{\pi^2(1+\lambda)^{5\over 2}\over 45}T^4+{(\sqrt{2} -1)(1+\lambda)\over 4\pi}{(eB)^{3\over2}T}\zeta_R\left(-{\textstyle\frac{1}{2}} \right)-{(1+\lambda)^{1\over 2}\over 48\pi^2}\left({eB}\right)^2\left(\gamma_E-{\lambda\over 2}+\ln{\sqrt{eB}\over 4\pi T}\right)\right],
\label{EU_1a}
\end{equation}
where the leading magnetic correction term is, overall, positive and produces attraction of the plates while the dominant Stefan Boltzmann term produces repulsion. Notice also that the magnetic correction term listed first is dominant when compared to the next.
The zero magnetic field limit of $E_U$ is most easily obtained by setting $F(eBt)=1$ in Eq. (\ref{zetaU0}) of Appendix I. After some straightforward manipulations, I obtain
\begin{equation}
E_U=
-L^2a\left[{\pi^2(1+\lambda)^{5\over 2}\over 45}T^4-{(1+\lambda)^{3\over 2}\over 12}M^2T^2+{(1+\lambda)\over 6\pi}M^3T+{(1+\lambda)^{1\over 2}\over 16\pi^2}{M^4}\left(\gamma_E-{\lambda\over 2}+\ln{M\over 4\pi T}\right)\right],
\label{EU_1b}
\end{equation}
where the leading order mass correction produces plates attraction.

In Appendix I, I show the details of how I obtain $E_{T, a}$ from $\zeta_{T, a}$, and find
\begin{equation}
E_{T,a}=E_a-(1+\lambda)^{\frac{3}{2}}{L^2T^2\over a}e^{-{4\pi  a T \sqrt{1+\lambda}}}e^{-{M^2a\over 2\pi T \sqrt{1+\lambda}}}F\left({eBa\over 2\pi T \sqrt{1+\lambda}}\right),
\label{ETa1}
\end{equation}
where $E_a$ is given by Eq. (\ref{Ea1}) of Appendix I when $a^{-1}\gg\sqrt{eB}, M$, by Eq. (\ref{Ea2}) when $\sqrt{eB}\gg a^{-1}, M$, and by Eq. (\ref{Ea2a}) when $M\gg a^{-1}, \sqrt{eB}$. As expected, the Casimir energy in the high temperature regime, $E=E_U+E_{T,a}$, does not contain
vacuum terms. Its leading order term is the Stefan Boltzmann term contained in the uniform energy density piece $E_U$. Notice also that the effects of mass and magnetic field are to produce plates attraction that reduce
the Stefan Boltzmann repulsion, as I will show in detail in Sec. \ref{5}.

The asymptotic limits of strong magnetic field, $\sqrt{eB}\gg a^{-1}, T, M$, and large mass, $M\gg a^{-1}, T, \sqrt{eB}$ are handled in a similar manner. I do a Poisson resummation on both the $m$-sum and the $n$-sum in Eq. (\ref{zeta1_2}), write the part dependent on $B$ as an infinite sum using Eq. (\ref{sinh}), and obtain
\begin{equation}
\zeta(s)={\mu^{2s}\sqrt{1+\lambda}\over\Gamma(s)}{L^2a\over 8\pi^2 T}\int_0^\infty dt\, t^{s-2}{e^{-M^2t}}eB\sum_{l=0}^\infty  e^{-(2l+1)eBt}\sum_{m=-\infty}^\infty e^{-{m^2\over 4T^2(1+\lambda)t}}
\sum_{n=-\infty}^\infty e^{-{n^2a^2\over t}}.
\label{zetaB_1}
\end{equation}
The term with $m=n=0$ is
\begin{equation}
\zeta_W(s)={\mu^{2s}\sqrt{1+\lambda}\over\Gamma(s)}{L^2a\over 8\pi^2 T}\int_0^\infty dt\, t^{s-2}{e^{-M^2t}}eB\sum_{l=0}^\infty  e^{-(2l+1)eBt},
\label{zetaW}
\end{equation}
the zeta function of the one-loop Weisskopf effective action for scalar QED, multiplied by the factor of $\sqrt{1+\lambda}$ that reflects the timelike asymmetry. Its contribution to the Casimir energy, $E_W=-T\zeta_W'(0)$, is a uniform energy density term independent of $T$. It is obtained using the same method I use in Appendix I to evaluate $\zeta_{U,0}(s)$, and yields the well known result
\begin{equation}
E_W={\sqrt{1+\lambda}L^2a\over 32\pi^2}e^2B^2\left[\left({1\over 3}-{M^4\over e^2B^2}\right)\left(1+\ln{{M^2\over 2eB}}\right)+8\zeta_H'(-1;{\textstyle\frac{1}{2}}+{\textstyle\frac{M^2}{2eB}})\right]
\label{EW_1}
\end{equation}
where $\zeta_H(s;z)$ is the Hurwitz zeta function defined in Appendix I, its derivative is
\begin{equation}
\zeta_H'(s;z)={\partial\zeta_H(s;z)\over\partial s},
\end{equation}
and I used
\begin{equation}
\zeta_H(-1;{\textstyle\frac{1}{2}}+{\textstyle\frac{z}{2}})= {1\over 24}-{z^2\over8}.
\end{equation}
Notice that I made the obvious choice $\mu=M$, indicating on-shell renormalization. I evaluate the rest of the zeta function by changing integration variable from $t$ to ${\sqrt{n^2a^2+{m^2\beta^2\over 4(1+\lambda)}}\over\sqrt{(2\ell+1)eB+M^2}}t$. Only terms with $n=\pm 1, m=0,\ell = 0$ and $n= 0,m=\pm 1,\ell = 0$ contribute significantly and I integrate them using the saddle point method. The Casimir energy in the strong magnetic field limit is then obtained using, once again, $E=-T\zeta'(0)$
\begin{equation}
E=E_W-{\sqrt{1+\lambda}L^2a\over 4\pi^{3/2}}eB\left(eB+M^2\right)^{\textstyle\frac{1}{4}}
\left[{e^{-2a\sqrt{eB+M^2}}\over a^{3/2}}+{\left(2\sqrt{1+\lambda}T\right)^{\textstyle\frac{3}{2}}}e^{-{\sqrt{eB+M^2}\over T\sqrt{1+\lambda}}}\right],
\label{EB_1}
\end{equation}
where $E_W$ in the strong magnetic field limit is 
\begin{equation}
E_W={\sqrt{1+\lambda}L^2a\over 32\pi^2}e^2B^2\left[{1\over 3}\left(1-\ln {4eB\over M^2} \right)-4\zeta_R'(-1)- {M^2\over eB}{\ln(4})-{M^4\over  e^2B^2}\left(\gamma_E-\ln {eB\over 2 M^2}\right)  \right],
\label{EW_2}
\end{equation}
obtained using the following identity and approximation \cite{Whittaker:1963xxy}
\begin{equation}
\zeta_H'(-1;{\textstyle\frac{1}{2}}+z)= -{1\over 2}\zeta_R'(-1) - {\ln 2\over 24} - {z\over 2}\ln(2\pi) +{z^2\over 2} +\int_0^z\ln\Gamma(x+{\textstyle\frac{1}{2}}) dx,
\end{equation}
\begin{equation}
\ln\Gamma(x+{\textstyle\frac{1}{2}}) \simeq {\ln \pi\over 2}-\left(\gamma_E+2\ln2\right)x+{\cal O}(x^2).
\end{equation}
The Casimir energy in the strong magnetic field approximation contains three parts: a uniform energy density, $E_W$, an exponentially suppressed vacuum part that I had already obtained in Ref. \cite{Erdas:2020ilo}, and the exponentially suppressed thermal correction. The uniform energy density will only contribute to the Casimir energy if the magnetic field is present between the plates and absent outside the plates. This term should not be considered if a uniform magnetic field is present in between and outside the plates.

Last I explore the large mass limit, $M\gg a^{-1}, T, \sqrt{eB}$. After a Poisson resummation of the $m$- and $n$-sums  in Eq. (\ref{zeta1_2}) I can, again, separate the term with $n=m=0$ from the rest, and that term is $\zeta_W(s)$ of Eq. (\ref{zetaW}). I evaluate the remaining part of that sum by changing variable of integration from $t$ to ${t\over M}\sqrt{n^2 a^2+{m^2\over 4T^2(1+\lambda)}}$. Only the terms with $m=\pm 1, n=0$ and $m=0, n=\pm 1$ contribute significantly to the sum, and I integrate them using the saddle point method. I then use $E=-T\zeta'(0)$ and find the Casimir energy in the large mass limit
\begin{equation}
E=E_W-{\sqrt{1+\lambda}L^2\over 8}\left({M\over\pi}\right)^{\textstyle\frac{3}{2}}
\left[{e^{-2Ma}\over a^{3/2}}F\left({eBa\over M}\right)+a{\left(2\sqrt{1+\lambda}T\right)^{\textstyle\frac{5}{2}}}e^{-{M\over T\sqrt{1+\lambda}}}F\left({eB\over 2MT\sqrt{1+\lambda}}\right)\right],
\label{EM_1}
\end{equation}
where
\begin{equation}
E_W=-{7\sqrt{1+\lambda}L^2a\over 5,760\pi^2}{e^4B^4\over M^4}, 
\label{EW_3}
\end{equation}
is the large mass limit of $E_W$ introduced in Eq. (\ref{EW_1}). Notice that I subtracted from $E_W$ a constant term proportional to $M^4$ that will not contribute to the Casimir energy. The large mass limit of the Casimir energy contains a uniform energy density part, $E_W$, an exponentially suppressed vacuum part in agreement with the one already obtained in my previous paper \cite{Erdas:2020ilo}, and an exponentially suppressed thermal correction.

\subsection{Spacelike anisotropy in the $x$-$y$ plane}
\label{3_3}
Using the eigenvalues (\ref{eigenvalues2}) and proceeding as I did when deriving Eq. (\ref{zeta1_1}), I find the zeta function for Lorentz anisotropy in the $x$-$y$ plane
\begin{equation}
\zeta(s)=\mu^{2s}\sum_{m=-\infty}^\infty \sum_{n=1}^\infty {L^2eB\over 2\pi}\sum_{\ell=0}^\infty \left[\left({2\pi m\over \beta}\right)^2+\left(1-{\lambda\over 2}\right)(2\ell+1)eB+\left({n\pi\over a}\right)^2+M^2\right]^{-s}.
\label{zeta2_1}
\end{equation}
A comparison of the zeta function for timelike anisotropy of Eq. (\ref{zeta1_1}) to that for spacelike anisotropy in the $x$-$y$ plane of Eq. (\ref{zeta2_1}), shows that (\ref{zeta2_1}) is obtained by setting $\lambda = 0 $ in
(\ref{zeta1_1}), then replacing $eB$ with $\left(1-{\lambda\over 2}\right)eB$ and finally dividing by $\left(1-{\lambda\over 2}\right)$. The same replacements can be applied to the vacuum and thermal parts of the Casimir energy obtained in the previous subsection and will produce those quantities in the case of spacelike anisotropy in the $x$-$y$ plane.
Notice that, for $\lambda\ll 1$, $\left(1-{\lambda\over 2}\right)\simeq \sqrt{1-\lambda}$ and therefore, from now on, $\left(1-{\lambda\over 2}\right)$ will be replaced with $\sqrt{1-\lambda}$. 

Making the replacements outlined above in Eqs. (\ref{E0_2}) and (\ref{ET_1}), I find the vacuum Casimir energy and its thermal correction in the limit of small plate distance
\begin{equation}
E_0=-{\pi^2L^2\over 8a^3\sqrt{1-\lambda}}\left[{1\over 90}-{M^2a^2\over 6\pi^2}+\left({M^4\over 2}-{(1-\lambda)e^2B^2\over 6}\right)\left({a\over \pi}\right)^4\left(\gamma_E  +\ln{\mu a\over 2\pi}\right)\right],
\label{E0_3}
\end{equation}
\begin{equation}
E_T=-{L^2T^2\over a\sqrt{1-\lambda}}e^{-{\pi\over a T }}e^{-{M^2a\over 2\pi T }}F\left({\sqrt{1-\lambda}eBa\over 2\pi T }\right),
\label{ET_2}
\end{equation}
where $\mu$ is defined below Eq. (\ref{E0_2}). I already obtained Eq. (\ref{E0_3}) in Ref. \cite{Erdas:2020ilo}, while Eq. (\ref{ET_2}) is a new result. Notice that in the absence of a magnetic field the vacuum energy for timelike anisotropy and the vacuum energy for spacelike anisotropy in the $x$-$y$ plane are identical since, for $\lambda \ll 1$, $\sqrt{1+\lambda}\simeq 1/\sqrt{1-\lambda}$, making this study that includes magnetic effects more compelling.

Moving to the case of high temperature, I make the replacements indicated above to Eqs. (\ref{EU_1}) and (\ref{ETa1}) and find the uniform energy density part and other part of the Casimir energy in the high temperature limit
\begin{equation}
E_U
=-{L^2a\over\sqrt{1-\lambda}} \left[{\pi^2T^4\over 45}-{M^2T^2\over 12}-{(1-\lambda)^{3\over 4}\over 2^{3\over 2}\pi}{(eB)^{3\over2}T}\zeta_H\left(-{\textstyle\frac{1}{2}} ; {\textstyle\frac{1}{2}}+{\textstyle\frac{M^2}{2\sqrt{1-\lambda}eB}}\right)+{{M^4}-(1-\lambda){e^2B^2\over 3}\over 16\pi^2}\left(\gamma_E+\ln{\mu\over 4\pi T}\right)\right],
\label{EU_2}
\end{equation}
\begin{equation}
E_{T,a}=E_a-{L^2T^2\over \sqrt{1-\lambda}a}e^{-{4\pi  a T}}e^{-{M^2a\over 2\pi T }}F\left({\sqrt{1-\lambda}eBa\over 2\pi T }\right),
\label{ETa2}
\end{equation}
where
\begin{equation}
E_{a}=-{L^2T\over 8\pi \sqrt{1-\lambda}a^2}\left[\zeta_R(3)+2M^2a^2\ln (2M a)+\left({M^4\over 6}-{(1-\lambda)e^2B^2\over 18}\right)a^4
\right]
\label{Ea3}
\end{equation}
when $a^{-1}\gg\sqrt{eB}, M$, 
\begin{equation}
E_{a}=-{L^2TeB\over 2\pi} e^{-2a\sqrt{M^2+\sqrt{1-\lambda}eB}}
\label{Ea4}
\end{equation}
when $\sqrt{eB}\gg a^{-1}, M$, and
\begin{equation}
E_{a}=-{L^2TM\over 4\pi \sqrt{1-\lambda}a} e^{-2Ma}F\left({\sqrt{1-\lambda}eBa\over M}\right)
\label{Ea4a}
\end{equation}
when $M\gg a^{-1}, \sqrt{eB}$.

The strong magnetic field limit is obtained by making the aforementioned replacements in Eqs. (\ref{EB_1}) and (\ref{EW_2})
\begin{equation}
E=E_W-{L^2a\over 4\pi^{3/2}}eB\left(\sqrt{1-\lambda}eB+M^2\right)^{\textstyle\frac{1}{4}}
\left[{e^{-2a\sqrt{\sqrt{1-\lambda}eB+M^2}}\over a^{3/2}}+{\left(2T\right)^{\textstyle\frac{3}{2}}}e^{-{\sqrt{\sqrt{1-\lambda}eB+M^2}\over T}}\right],
\label{EB_2}
\end{equation}
where $E_W$ in the strong magnetic field limit is 
\begin{equation}
E_W={L^2a\over 32\pi^2}\sqrt{1-\lambda}e^2B^2\left[{1\over 3}\left(1+{\lambda\over 2}-\ln{{4eB}\over{M^2}}\right)-4\zeta_R'(-1)-{M^2\over \sqrt{1-\lambda}eB}{\ln(4})-{M^4\over  (1-\lambda)e^2B^2}\left(\gamma_E+{\lambda\over 2}-\ln {eB\over 2 M^2}\right)\right].
\label{EW_4}
\end{equation}

The large mass limit is obtained in the same manner, using Eqs. (\ref{EM_1}) and (\ref{EW_3})
\begin{equation}
E=E_W-{L^2\over 8 \sqrt{1-\lambda} }\left({M\over\pi}\right)^{\textstyle\frac{3}{2}}
\left[{e^{-2aM}\over a^{3/2}}F\left({\sqrt{1-\lambda}eBa\over M}\right)+a{\left(2T\right)^{\textstyle\frac{5}{2}}}e^{-{M\over T}}F\left({\sqrt{1-\lambda}eB\over 2MT}\right)\right],
\label{EM_2}
\end{equation}
with
\begin{equation}
E_W=-{7L^2a\over 5,760\pi^2}{(1-\lambda)^{\frac{3}{2}}e^4B^4\over M^4}.
\label{EW_5}
\end{equation}


\subsection{Spacelike anisotropy in the $z$ direction}
\label{3_4}
I obtain the zeta function in the case of a spacelike anisotropy in the $z$ direction using the eigenvalues (\ref{eigenvalues3})
\begin{equation}
\zeta(s)=\mu^{2s}\sum_{m=-\infty}^\infty \sum_{n=1}^\infty {L^2eB\over 2\pi}\sum_{\ell=0}^\infty \left[\left({2\pi m\over \beta}\right)^2+(2\ell+1)eB+\left(1-{\lambda}\right)\left({n\pi\over a}\right)^2+M^2\right]^{-s}.
\label{zeta3_1}
\end{equation}
When comparing the zeta function for timelike anisotropy of Eq. (\ref{zeta1_1}) to that for spacelike anisotropy in the $z$ direction of Eq. (\ref{zeta3_1}), it is evident that (\ref{zeta3_1}) is obtained by setting $\lambda = 0 $ in
(\ref{zeta1_1}), then replacing $a$ with $a\over \sqrt{1-\lambda}$. The same replacements, when applied to the vacuum and thermal parts of the Casimir energy obtained in subsection \ref{3_2}, yield those quantities in the case of spacelike anisotropy in the $z$ direction.

Making those replacements in Eqs. (\ref{E0_2}) and (\ref{ET_1}), I obtain the vacuum Casimir energy and its thermal correction in the limit of small plate distance
\begin{equation}
E_0=-{\pi^2L^2(1-\lambda)^{3\over 2}\over 8a^3}\left[{1\over 90}-{M^2a^2\over 6\pi^2(1-\lambda)}+\left({M^4\over 2}-{e^2B^2\over 6}\right)\left({a^4\over \pi^4(1-\lambda)^2}\right)\left(\gamma_E  +\ln{\mu a\over 2\pi}+{\lambda\over 2}\right)\right],
\label{E0_4}
\end{equation}
\begin{equation}
E_T=-{L^2T^2\sqrt{1-\lambda}\over a}e^{-{\pi\sqrt{1-\lambda}\over a T }}e^{-{M^2a\over 2\pi T\sqrt{1-\lambda} }}F\left({eBa\over 2\pi \sqrt{1-\lambda}T }\right),
\label{ET_3}
\end{equation}
with $\mu$ defined below Eq. (\ref{E0_2}). I already obtained Eq. (\ref{E0_4}) in Ref. \cite{Erdas:2020ilo}, while its thermal correction of Eq. (\ref{ET_3}) is a new result.

In the limit of high temperature, I find
\begin{equation}
E_U
=-{L^2a\over\sqrt{1-\lambda}} \left[{\pi^2T^4\over 45}-{M^2T^2\over 12}-{1\over 2^{3\over 2}\pi}{(eB)^{3\over2}T}\zeta_H\left(-{\textstyle\frac{1}{2}} ; {\textstyle\frac{1}{2}}+{\textstyle\frac{M^2}{2eB}}\right)+{{M^4}-{e^2B^2\over 3}\over 16\pi^2}\left(\gamma_E+\ln{\mu\over 4\pi T}\right)\right],
\label{EU_3}
\end{equation}
\begin{equation}
E_{T,a}=E_a-{L^2 \sqrt{1-\lambda} T^2\over a}e^{-{4\pi  a T\over\sqrt{1-\lambda}}}e^{-{M^2a\over 2\pi \sqrt{1-\lambda}T }}F\left({eBa\over 2\pi \sqrt{1-\lambda}T }\right),
\label{ETa3}
\end{equation}
where
\begin{equation}
E_{a}=-{L^2(1-\lambda)T\over 8\pi a^2}\left[\zeta_R(3)+{2M^2a^2\over(1-\lambda)}\ln \left({2M a\over\sqrt{1-\lambda}}\right)+\left({M^4\over 6}-{e^2B^2\over 18}\right){a^4\over(1-\lambda)^2}
\right]
\label{Ea5}
\end{equation}
when $a^{-1}\gg\sqrt{eB}, M$,
\begin{equation}
E_{a}=-{L^2TeB\over 2\pi} e^{-2a{\sqrt{eB+M^2\over 1-\lambda} }}
\label{Ea6}
\end{equation}
when $\sqrt{eB}\gg, a^{-1} M$, and 
\begin{equation}
E_{a}=-{L^2\sqrt{1-\lambda}TM\over 4\pi a} e^{-2a M\over \sqrt{1-\lambda}}F\left({eBa\over \sqrt{1-\lambda}M}\right)
\label{Ea6a}
\end{equation}
when $M\gg, a^{-1} \sqrt{eB}$.

I find the strong magnetic field limit of the Casimir energy in the case of anisotropy in the $z$ direction by making the replacements in Eqs. (\ref{EB_1}) and (\ref{EW_2})
\begin{equation}
E=E_W-{L^2a\over 4\pi^{3/2}\sqrt{1-\lambda}}eB\left(eB+M^2\right)^{\textstyle\frac{1}{4}}
\left[\left({\sqrt{1-\lambda}\over a}\right)^{\textstyle\frac{3}{2}}{e^{-2a{\sqrt{eB+M^2\over1-\lambda}}}}+{\left(2T\right)^{\textstyle\frac{3}{2}}}e^{-{\sqrt{eB+M^2}\over T}}\right],
\label{EB_3}
\end{equation}
where $E_W$ is 
\begin{equation}
E_W={L^2a\over 32\pi^2 \sqrt{1-\lambda}}e^2B^2\left[{1\over 3}\left(1-\ln{4eB\over M^2}\right)-4\zeta_R'(-1)-{M^2\over eB}{\ln(4})-\left(\gamma_E-{M^4\over e^2B^2}\ln {eB\over 2 M^2}\right)\right].
\label{EW_6}
\end{equation}

I obtain the large mass limit in the same manner using Eqs. (\ref{EM_1}) and (\ref{EW_3})
\begin{equation}
E=E_W-{L^2\over 8  }\left({M\over\pi}\right)^{\textstyle\frac{3}{2}}
\left[\left({\sqrt{1-\lambda}\over a}\right)^{\textstyle\frac{3}{2}}{e^{-{2Ma\over  \sqrt{1-\lambda}}}}F\left({eBa\over M \sqrt{1-\lambda}}\right)+{a\over \sqrt{1-\lambda}}{\left(2T\right)^{\textstyle\frac{5}{2}}}e^{-{M\over T}}F\left({eB\over 2MT}\right)\right],
\label{EM_3}
\end{equation}
with
\begin{equation}
E_W=-{7L^2a\over 5,760\pi^2\sqrt{1-\lambda}}{e^4B^4\over M^4},
\label{EW_7}
\end{equation}
where the vacuum part was obtained in my previous work, Ref. \cite{Erdas:2020ilo}, and its thermal correction is another new result.

\section{Mixed boundary conditions}
\label{4}
Mixed boundary conditions constrain the scalar field
to vanish at the plate located at $z=0$
\begin{equation}
\phi(x, y, 0,\tau)=0,
\label{mixed1}
\end{equation}
and its normal derivative to vanish at the plate located at $z=a$ ,
\begin{equation}
{\partial\phi\over\partial z}(x, y, a,\tau)=0,
\label{mixed2}
\end{equation}
yielding
\begin{equation}
k_z=\left(n+\textstyle{\frac{1}{2}}\right){\pi\over a},
\label{mixed3}
\end{equation}
with $n=0,1,2,\cdots$.

The zeta function for timelike asymmetry and mixed boundary conditions is constructed with the same method used to obtain the one for timelike asymmetry and Dirichlet boundary conditions of Eq. (\ref{zeta1_1}), the only difference being that $k_z$ is given by Eq. (\ref{mixed3})
\begin{equation}
\zeta(s)=\mu^{2s}(1+\lambda)\sum_{m=-\infty}^\infty \sum_{n=0}^\infty {L^2eB\over 2\pi}\sum_{\ell=0}^\infty \left[(1+\lambda)\left({2\pi m\over \beta}\right)^2+(2\ell+1)eB+\left(n+{1\over 2}\right)^2
\left({\pi\over a}\right)^2+M^2\right]^{-s}.
\label{zetam_1}
\end{equation}
The zeta function for mixed boundary conditions and spacelike asymmetry in the $x$-$y$ plane is quickly obtained from Eq. (\ref{zeta2_1}) by replacing $n$ with $n+{1\over2}$. The same replacement made in Eq. (\ref{zeta3_1}) produces the 
zeta function for mixed boundary conditions and spacelike asymmetry in the $z$ direction.

Using identities (\ref{z-s}) and (\ref{sinh}) I write (\ref{zetam_1}) as
\begin{equation}
\zeta(s)={\mu^{2s}(1+\lambda)\over\Gamma(s)}{L^2eB\over 4\pi}\int_0^\infty dt\, t^{s-1}{e^{-M^2t}\over\sinh eBt}\sum_{m=-\infty}^\infty  e^{-(1+\lambda){4\pi^2 m^2\over \beta^2}t}\sum_{n=0}^\infty e^{-\left(n+{1\over2}\right)^2{\pi^2\over a^2}t}.
\label{zetam_2}
\end{equation}

To investigate the asymptotic case of small plate distance and timelike asymmetry, I apply the Poisson resummation method to the $m$-sum, as I did in Eq. (\ref{Poisson1}), and separate the zeta function (\ref{zetam_2}) into a vacuum part and a thermal part.
The vacuum part of the zeta function yields the vacuum Casimir energy I already obtained in Ref. \cite{Erdas:2020ilo}
\begin{equation}
E_0={\pi^2\sqrt{1+\lambda}L^2\over 8a^3}\left[{7\over 720}-{M^2a^2\over 12\pi^2}-\left({M^4\over 2}-{e^2B^2\over 6}\right)\left({a\over \pi}\right)^4\left(\gamma_E  +\ln{2\mu a\over \pi}\right)\right].
\label{Em0_2}
\end{equation}
The thermal part is handled by making a change of integration variable from $t$ to ${ma\over 2\pi \left(n+{1\over 2}\right) T\sqrt{1+\lambda}}t$. In the limit of small plate distance only the terms with $n=0$
and $m=\pm 1$ contribute significantly, I evaluate the remaining integral using the saddle point method and obtain the following thermal part of the Casimir energy for timelike asymmetry and mixed boundary conditions in the small plate distance limit
\begin{equation}
E_T=-{(1+\lambda)^{3/2}L^2T^2\over 2a}e^{-{\pi\over 2a T \sqrt{1+\lambda}}}e^{-{M^2a\over \pi T \sqrt{1+\lambda}}}F\left({eBa\over \pi T \sqrt{1+\lambda}}\right).
\label{EmT_1}
\end{equation}
Notice that, in the small plate distance limit, the vacuum Casimir energy is negative for Dirichlet boundary conditions and positive for mixed boundary conditions, while the thermal correction is the same for both Dirichlet and mixed boundary conditions. As I will show in detail in Sec. \ref{5}, thermal effects decrease the plates attraction in the case of Dirichlet boundary conditions and increase the plates repulsion in the case of mixed boundary conditions.

 The vacuum and thermal parts of the Casimir energy for spacelike asymmetry in the $x$-$y$ plane in the small plate distance limit are obtained from Eqs. (\ref{Em0_2}) and (\ref{EmT_1}) by setting $\lambda = 0 $ in
them, then replacing $eB$ with $\left(1-{\lambda\over 2}\right)eB$ and finally dividing by $\left(1-{\lambda\over 2}\right)$. For the sake of brevity I will not report those two quantities.
Notice that in the absence of a magnetic field the vacuum energy for timelike anisotropy and the vacuum energy for spacelike anisotropy in the $x$-$y$ plane are indistinguishable, as it was the case when Dirichlet boundary conditions were investigated.

$E_0$ and $E_T$ for spacelike asymmetry in the $z$ direction in the small plate distance limit are obtained from Eqs. (\ref{Em0_2}) and (\ref{EmT_1}) by setting $\lambda = 0 $ in
them, then replacing $a$ with $a\over\sqrt{1-\lambda}$.

I obtain the high temperature limit of the Casimir energy for mixed boundary conditions and timelike anisotropy by doing a Poisson resummation of the $n$-sum in Eq. (\ref{zetam_2})
\begin{equation}
\sum_{n=0}^\infty e^{-{(n+{\frac{1}{2}})^2\pi^2\over a^2}t} = {a\over 2\sqrt{\pi t}}\sum_{n=-\infty}^\infty (-1)^ne^{-{n^2a^2\over t}},
\label{Poisson3}
\end{equation}
and separate the zeta function into a uniform density part, $\zeta_U(s)$, and a remaining part, $\zeta_{T,a}(s)$, as I did when investigating Dirichlet boundary conditions. The uniform energy density piece of the Casimir energy, $E_U$, is the same as in the case of Dirichlet boundary conditions, Eq. (\ref{EU_1}). The other part of the Casimir energy is
\begin{equation}
E_{T,a}=E_a+(1+\lambda)^{\frac{3}{2}}{L^2T^2\over a}e^{-{4\pi  a T \sqrt{1+\lambda}}}e^{-{M^2a\over 2\pi T \sqrt{1+\lambda}}}F\left({eBa\over 2\pi T \sqrt{1+\lambda}}\right),
\label{ETa4}
\end{equation}
with
\begin{equation}
E_{a}={(1+\lambda)}{L^2T\over 4\pi a^2}\left[{3\over 8}\zeta_R(3)-\ln(2)M^2a^2
-\left({M^4\over 4}-{e^2B^2\over 12}\right)a^4
\right],
\label{Ea7}
\end{equation}
for $a^{-1}\gg\sqrt{eB}, M$, 
\begin{equation}
E_{a}={(1+\lambda)L^2TeB\over 2\pi} e^{-2a\sqrt{M^2+eB}}
\label{Ea8}
\end{equation}
for $\sqrt{eB}\gg a^{-1}, M$, and
\begin{equation}
E_{a}={(1+\lambda)L^2TM\over 4\pi a} e^{-2aM}F\left({eBa\over M}\right)
\label{Ea8a}
\end{equation}
for $M\gg a^{-1}, \sqrt{eB}$.
The details of this calculation are shown in Appendix II. Notice that the dominant part of the Casimir energy is the uniform energy density piece, with the Stefan Boltzmann term, as it was the case when Dirichlet boundary conditions were considered.
The high temperature Casimir energy for mixed boundary conditions and spacelike asymmetry in the $x$-$y$ plane is obtained from Eqs. (\ref{EU_1}), (\ref{ETa4}), (\ref{Ea7}) and (\ref{Ea8}) by setting $\lambda=0$, dividing by
$\sqrt{1-\lambda}$ and replacing $eB$ with $\sqrt{1-\lambda}eB$, as described previously. The Casimir energy for the case of spacelike asymmetry in the $z$ direction is obtained from the four equations cited above by setting $\lambda=0$ and then replacing $a$ with $a\over\sqrt{1-\lambda}$.

I obtain the mixed boundary conditions Casimir energy for strong magnetic field, in the case of timelike Lorentz asymmetry, by doing a Poisson resummation of the $m$- and $n$-series in Eq. (\ref{zetam_2}) and by writing the $B$-dependent part as an infinite sum, to find
\begin{equation}
\zeta(s)={\mu^{2s}\sqrt{1+\lambda}\over\Gamma(s)}{L^2a\over 8\pi^2 T}\int_0^\infty dt\, t^{s-2}{e^{-M^2t}}eB\sum_{l=0}^\infty  e^{-(2l+1)eBt}\sum_{m=-\infty}^\infty e^{-{m^2\over 4T^2(1+\lambda)t}}
\sum_{n=-\infty}^\infty (-1)^n e^{-{n^2a^2\over t}}.
\label{zetaBm_1}
\end{equation}
At this point I separate the term with $m=n=0$ that yields the Weisskopf zeta function, $\zeta_W(s)$, already examined in Sec. \ref{3}, which contributes $E_W$ of Eq. (\ref{EW_2}) to the Casimir energy. The remaining terms in the sum are handled by a change of integration variable from $t$ to ${\sqrt{n^2a^2+{m^2\beta^2\over 4(1+\lambda)}}\over\sqrt{(2\ell+1)eB+M^2}}t$, as I did in Sec. \ref{3} for Dirichlet boundary conditions, that produces the same outcome as in Sec. \ref{3}, leaving only terms with $n=\pm1, m=0, \ell=0$ and $n=0, m= \pm 1, \ell=0$ to contribute significantly to the sum. The Casimir energy I obtain is
\begin{equation}
E=E_W+{\sqrt{1+\lambda}L^2a\over 4\pi^{3/2}}eB\left(eB+M^2\right)^{\textstyle\frac{1}{4}}
\left[{e^{-2a\sqrt{eB+M^2}}\over a^{3/2}}-{\left(2\sqrt{1+\lambda}T\right)^{\textstyle\frac{3}{2}}}e^{-{\sqrt{eB+M^2}\over T\sqrt{1+\lambda}}}\right],
\label{EBm_1}
\end{equation}
where the vacuum part has opposite sign to that obtained for Dirichlet boundary conditions, agreeing with what I found in Ref. \cite{Erdas:2020ilo}. The thermal correction to the Casimir energy in the strong magnetic field limit is the same for mixed and Dirichlet boundary conditions. The strong magnetic field Casimir energy for mixed boundary conditions and Lorentz asymmetry in the $x$-$y$ plane is the same as that for Dirichlet boundary conditions with a factor of $-1$ that will multiply the vacuum part. The one for Lorentz asymmetry in the $z$ direction is also the same as that for Dirichlet boundary with the vacuum part having opposite sign.

The last asymptotic case I examine is the one of large mass for mixed boundary conditions. Starting with timelike Lorentz asymmetry, I proceed as I did above by doing a Poisson resummation of the $m$- and $n$-series in Eq. (\ref{zetam_2}). The term with $m=n=0$ is the Weisskopf zeta function which, in the large mass limit, contributes $E_W$ of Eq. (\ref{EW_3}) to the Casimir energy. I evaluate the rest of the sum by doing a change of integration variable from $t$ to ${t\over M}{\sqrt{n^2a^2+{m^2\beta^2\over 4(1+\lambda)}}}$, retain only the significant terms with $n=\pm1, m=0$ and $n=0, m=\pm1$ and obtain the following for the Casimir energy
\begin{equation}
E=E_W+{\sqrt{1+\lambda}L^2\over 8}\left({M\over\pi}\right)^{\textstyle\frac{3}{2}}
\left[{e^{-2aM}\over a^{3/2}}F\left({eBa\over M}\right)-a{\left(2\sqrt{1+\lambda}T\right)^{\textstyle\frac{5}{2}}}e^{-{M\over T\sqrt{1+\lambda}}}F\left({eB\over 2MT\sqrt{1+\lambda}}\right)\right],
\label{EMm_1}
\end{equation}
where, again, the vacuum part is equal and opposite to the vacuum part obtained for Dirichlet boundary conditions, while the thermal correction and $E_W$ are the same as in the case of Dirichlet boundary conditions. The vacuum part agrees with my result of Ref. \cite{Erdas:2020ilo}. The large mass Casimir energy for mixed boundary conditions and Lorentz asymmetry in the $x$-$y$ plane has the same thermal piece and $E_W$ and opposite vacuum part as that I obtained for Dirichlet boundary conditions, and the same is true for the case of Lorentz asymmetry in the $z$-direction.

\section{Casimir pressure}
\label{5}
In this section I will examine the Casimir pressure for all asymptotic cases and timelike anisotropy, with either Dirichlet or mixed boundary conditions. Once the pressure for a certain case of timelike anisotropy is known, the pressure for the two corresponding cases of spacelike anisotropy is obtained immediately with the simple substitutions outlined in the previous sections.

I obtain the Casimir pressure, $P$, by taking a derivative of the Casimir energy
\begin{equation}
P=-{1\over L^2}{\partial E\over\partial a}.
\label{P_1}
\end{equation}

I first obtain the Casimir pressure for timelike Lorentz asymmetry and Dirichlet boundary in the asymptotic case of small plate distance,
\begin{equation}
P=-{\pi^2\sqrt{1+\lambda}\over 8a^4}\left[{1\over 30}-{M^2a^2\over 6\pi^2}+{e^2B^2\over 6}\left({a\over \pi}\right)^4\left(1+\gamma_E  +\ln{\mu a\over 2\pi}\right)\right]+{(1+\lambda)\pi T\over a^3}e^{-{\pi\over a T \sqrt{1+\lambda}}}e^{-{M^2a\over 2\pi T \sqrt{1+\lambda}}}F\left({eBa\over 2\pi T \sqrt{1+\lambda}}\right),
\label{P_2}
\end{equation}
where smaller terms have been neglected. Notice how the thermal component of the pressure is repulsive and exponentially suppressed, but will have a significant contribution when $0.5\ge aT \ge 0.1$, and how the vacuum magnetic term gives almost a constant contribution, aside from a weak logarithmic dependence on $a$. The Casimir pressure for the same asymptotic case but mixed boundary conditions is
\begin{equation}
P={\pi^2\sqrt{1+\lambda}\over 8a^4}\left[{7\over 240}-{M^2a^2\over 12\pi^2}-{e^2B^2\over 6}\left({a\over \pi}\right)^4\left(1+\gamma_E  +\ln{2\mu a\over \pi}\right)\right]+{(1+\lambda)\pi T\over a^3}e^{-{\pi\over a T \sqrt{1+\lambda}}}e^{-{M^2a\over 2\pi T \sqrt{1+\lambda}}}F\left({eBa\over 2\pi T \sqrt{1+\lambda}}\right),
\label{P_3}
\end{equation}
with its vacuum part displaying the well known repulsive nature and with the same thermal piece as in the case of Dirichlet boundary conditions. It is interesting how the vacuum magnetic term is almost identical, attractive and practically constant in both cases.

The high temperature limit of the Casimir pressure, in the presence of a timelike anisotropy and Dirichlet boundary conditions, is given by
\begin{eqnarray}
P&=&\!\!\!\
{\pi^2(1+\lambda)^{5\over 2}\over 45}T^4-{(1+\lambda)^{3\over 2}\over 12}M^2T^2+P_a-{(1+\lambda)\over 2^{3\over 2}\pi}{(eB)^{3\over2}T}\zeta_H\left(-{\textstyle\frac{1}{2}} ; {\textstyle\frac{1}{2}}+{\textstyle\frac{M^2}{2eB}}\right)
\nonumber \\
&-&\!\!\!\
{4\pi (1+\lambda)^2T^3\over a}e^{-4{\pi a T \sqrt{1+\lambda}}}e^{-{M^2a\over 2\pi T \sqrt{1+\lambda}}}F\left({eBa\over 2\pi T \sqrt{1+\lambda}}\right),
\label{P_4}
\end{eqnarray}
where I neglected several lower order terms and use the notation
\begin{equation}
P_a=-{1\over L^2}{\partial E_a\over\partial a},
\label{P_a}
\end{equation}
with $E_a$ defined in Eqs. (\ref{Ea1}), (\ref{Ea2}), and (\ref{Ea2a}) of Appendix I.  I find 
\begin{equation}
P_a=-{(1+\lambda) \over 4\pi a^3}T\zeta_R(3)
\label{P_a1}
\end{equation}
for $a^{-1}\gg\sqrt{eB}, M$,
\begin{equation}
P_a=-{(1+\lambda)\over \pi} TeB\sqrt{eB+M^2} e^{-2a\sqrt{eB+M^2}}
\label{P_a2}
\end{equation}
for $\sqrt{eB} \gg a^{-1}, M$, and
\begin{equation}
P_a=-{(1+\lambda)TM^2\over 2\pi a} e^{-2Ma}F\left({eBa\over M}\right)
\label{P_a3}
\end{equation}
for $M \gg a^{-1}, \sqrt{eB}$. Notice that I neglected some smaller terms in Eqs. (\ref{P_a1}) - (\ref{P_a3}). The contribution of $P_a$ to the Casimir pressure is negligibly small in all cases shown above, except when $a^{-1}\gg\sqrt{eB}, M$. The leading order term is the repulsive Stefan Boltzmann term, independent of the plate distance. All other terms, mass, magnetic field, and plate distance corrections, are attractive and "conspiring" to reduce the Stefan Boltzmann repulsion. 

The high temperature pressure for mixed boundary conditions and timelike anisotropy is 
\begin{eqnarray}
P&=&\!\!\!\
{\pi^2(1+\lambda)^{5\over 2}\over 45}T^4-{(1+\lambda)^{3\over 2}\over 12}M^2T^2+P_a-{(1+\lambda)\over 2^{3\over 2}\pi}{(eB)^{3\over2}T}\zeta_H\left(-{\textstyle\frac{1}{2}} ; {\textstyle\frac{1}{2}}+{\textstyle\frac{M^2}{2eB}}\right)
\nonumber \\
&+&\!\!\!\
{4\pi (1+\lambda)^2T^3\over a}e^{-4{\pi a T \sqrt{1+\lambda}}}e^{-{M^2a\over 2\pi T \sqrt{1+\lambda}}}F\left({eBa\over 2\pi T \sqrt{1+\lambda}}\right),
\label{P_5}
\end{eqnarray}
where
\begin{equation}
P_a={3(1+\lambda) \over 16\pi a^3}T\zeta_R(3)
\label{P_am}
\end{equation}
for $a^{-1}\gg\sqrt{eB}, M$, and $P_a$ is negligibly small in the other two cases. Under these conditions, the contribution of the uniform energy density part is the same as in the case of Dirichlet boundary, with the dominating, repulsive, Stefan Boltzmann term and the attractive mass and magnetic field corrections. The plate distance correction, $P_a$, is repulsive in this case as expected, and the exponentially decaying correction is repulsive too.

The Casimir pressure for strong magnetic field and Dirichlet boundary conditions, in the case of timelike anisotropy is given by
\begin{equation}
P={\sqrt{1+\lambda}\over 96\pi^2}e^2B^2\left(C+\ln {eB\over M^2} \right)-{\sqrt{1+\lambda}\over 2\pi^{3/2}}eB\left(eB+M^2\right)^{\textstyle\frac{1}{4}}
\left[\sqrt{eB+M^2\over a}{e^{-2a\sqrt{eB+M^2}}}-\sqrt{2}{\left(\sqrt{1+\lambda}T\right)^{\textstyle\frac{3}{2}}}e^{-{\sqrt{eB+M^2}\over T\sqrt{1+\lambda}}}\right],
\label{P_6}
\end{equation}
where $C=12\zeta_R'(-1)-1+\ln 4= -1.59876$, and where the dominant term is the Weisskopf term, repulsive when $eB> e^{-C} M^2$, with $e^{-C}\simeq 5$. The other vacuum term, exponentially suppressed, is attractive while the thermal correction, also exponentially suppressed, is repulsive. The pressure in the case of strong magnetic field and mixed boundary conditions is very similar to what I show in Eq. (\ref{P_6}), the only difference being that the exponentially suppressed vacuum term is equal but overall positive and therefore repulsive.

When investigating the case of large mass, Dirichlet boundary conditions, and timelike anisotropy, I find
\begin{equation}
P={7\sqrt{1+\lambda}\over 5,760\pi^2}{e^4B^4\over M^4}-{\sqrt{1+\lambda}\over 4}\left({M\over\pi}\right)^{\textstyle\frac{3}{2}}
\left[M{e^{-2aM}\over a^{3/2}}F\left({eBa\over M}\right)-2^{\textstyle\frac{3}{2}}{\left(\sqrt{1+\lambda}T\right)^{\textstyle\frac{5}{2}}}e^{-{M\over T\sqrt{1+\lambda}}}F\left({eB\over 2MT\sqrt{1+\lambda}}\right)\right],
\label{P_7}
\end{equation}
notice that also in this case the exponentially suppressed vacuum term produces attraction, while the thermal correction and Weisskopf term are repulsive, as always. This is the only asymptotic case where a dominant term is missing and therefore the pressure can be attractive or repulsive depending on the relative size of $B$, $a$, and $M$. Finally, the large mass pressure for mixed boundary conditions is identical to Eq. (\ref{P_7}), with the exception of the exponentially suppressed vacuum term that has opposite sign. A positive vacuum term causes the pressure to be certainly repulsive, regardless of the relative size of $B$, $a$, and $M$.
\section{Discussion and conclusions}
\label{6}

I used the zeta function regularization method to investigate the finite temperature Casimir effect due to a Lorentz-violating scalar field in the presence of a uniform magnetic field perpendicular to the plates. The scalar field satisfies a modified version of the Klein-Gordon equation, where the field derivative is coupled to a unit four-vector $u^\mu$ thus breaking Lorentz invariance in a CPT-even aether-like way. I investigated in detail Dirichlet and mixed boundary conditions. Neumann boundary conditions have not been considered since they produce the same results as Dirichlet.

I derived simple analytic expressions for the thermal corrections to the Casimir energy in the asymptotic case of small plate distance and Dirichlet boundary conditions when $u^\mu$ is timelike, spacelike and perpendicular to $\bf B$, and spacelike and parallel to $\bf B$, reporting my results in Eqs. (\ref{ET_1}), (\ref{ET_2}), and (\ref{ET_3}) respectively. I obtained the same quantities for the case of mixed boundary conditions, Eq. (\ref{EmT_1}). I obtained also simple analytic expressions for the thermal corrections to the Casimir energy in the asymptotic cases of strong magnetic field and large mass for the three types of space-time anisotropy listed above in the case of Dirichlet boundary conditions, Eqs. (\ref{EB_1}), (\ref{EB_2}), (\ref{EB_3}), and Eqs. (\ref{EM_1}), (\ref{EM_2}), (\ref{EM_3}), and in the case of mixed boundary conditions, Eqs. (\ref{EBm_1}), (\ref{EMm_1}). I derived the Casimir energy in the asymptotic case of high temperature for the three possible types of space-time anisotropy and Dirichlet boundary conditions, Eqs. (\ref{EU_1}),  (\ref{ETa1}), Eqs. (\ref{EU_2}),  (\ref{ETa2}), and Eqs. (\ref{EU_3}),  (\ref{ETa3}), and for mixed boundary conditions Eqs. (\ref{ETa4}). The Casimir energy for the cases of high temperature, strong magnetic field, and large mass contain a uniform energy density piece which is dominant in the high temperature case because of the Stefan-Boltzmann term, and in the strong magnetic field case with the Weisskopf term, regardless of the type of boundary condition considered. The uniform energy density terms are equal for both sets of boundary conditions considered. 
I find that the thermal correction to the Casimir energy and pressure in the asymptotic cases of small plate distance, strong magnetic field, and large mass, is exponentially suppressed for all cases of Lorentz asymmetry considered and for Dirichlet or mixed boundary conditions. I discover that the thermal correction to the pressure is repulsive in the asymptotic cases of small plate distance, strong magnetic field, and large mass, and for all cases of Lorentz asymmetry and boundary conditions considered.

My results show a strong dependence of the Casimir energy and pressure on the Lorentz symmetry breaking parameter $\lambda$, as observed by Refs. \cite{Cruz:2017kfo, Cruz:2018bqt,Erdas:2020ilo}. The vacuum parts of the results obtained for the cases of small plate distance and large mass agree to leading order with Ref. \cite{Cruz:2017kfo} when $B=0$, and agree with Ref. \cite{Erdas:2020ilo} when $B$ is considered. Notice that Refs. \cite{Cruz:2017kfo,Cruz:2018bqt} do not include a magnetic field in their study. The vacuum part of the Casimir energy and pressure obtained for the strong magnetic field case is also in full agreement with Ref. \cite{Erdas:2020ilo}, for the three types of Lorentz asymmetry considered. A comparison of the thermal parts of my results with Ref. \cite{Cruz:2018bqt} where a different regularization method is used, the Abel-Plana summation formula, is more difficult since most of their results are expressed as infinite sums, and only in a few specific cases the authors are able to obtain closed forms for their results. The asymptotic limits examined on my paper are more general than those considered in \cite{Cruz:2018bqt}. For example, the large mass limit I consider is valid for any values of $a^{-1}$, $T$, and $\sqrt{eB}$ as long as they are all smaller than $M$, and similarly for the other three asymptotic cases I consider. On the other hand, Ref. \cite{Cruz:2018bqt}, which does not include a magnetic field in their study, considers asymptotic cases such as $M\gg a^{-1}\gg T$ where the thermal part of their and mine results agree on the exponential suppression but disagree on less important factors of $M$ and $T$ that multiply the exponential. In the asymptotic case of high temperature, $T\gg a^{-1} \gg M$ for Ref. \cite{Cruz:2018bqt}, my result and that of Ref. \cite{Cruz:2018bqt} agree on the leading Stefan Boltzmann term having a $T^4$ dependence and on its dependence on $\lambda$, but disagree on the numerical factor multiplying this term. However, when taking $\lambda =0$ in the results I obtain for the high temperature Casimir energy, one obtains values that are in full agreement with the well known Stefan Boltzmann terms for isotropic space-time for either set of boundary conditions.

It is difficult to estimate what the value of $\lambda$ could be. While it is certain that $\lambda\ll 1$, Cruz et al. \cite{Cruz:2017kfo, Cruz:2018bqt} show numerical results evaluated for $\lambda = 0.1, 0.2$. If the Lorentz violating parameter were to lie in that range, the temperature and magnetic corrections I obtain here would be of the order of 5 - 15$\%$ of the value they have in the case of unbroken Lorentz symmetry. However, these estimates of the value of $\lambda$ are almost certainly too high. Several papers that examine Lorentz violation within QED appeared in the literature, some within the Casimir effect \cite{Frank:2006ww, Kharlanov:2009pv}, others within other contexts \cite{Klinkhamer:2010zs, Exirifard:2010xm, Casana:2011vh, Casana:2011du}.
A lower estimate of $\lambda$ was proposed in Ref. \cite{Kharlanov:2009pv} within the context of Lorentz violating extended QED, where the leading $\lambda$-correction to the Casimir pressure is about $1\% - 2\%$ of the value of the pressure in isotropic space-time. Another paper \cite{Frank:2006ww}, studying  extensions of QED from the Lorentz-violating scalar and fermion sector of the minimally extended Standard Model, reported an additive correction to the Casimir force due to CPT-odd Lorentz-violating terms that is consistent within 15$\%$ of experimental data. Other papers \cite{Klinkhamer:2010zs,Exirifard:2010xm}
that do not look at the Casimir effect but examine other direct experimental evidence or look at indirect bounds of Lorentz violation, estimate Lorentz-violating effects in QED to be around one part in $10^{8}$. While I am not investigating QED in this paper, it seems that, for the scalar theory I am considering, an estimate of $\lambda\le 0.01-0.02$ is reasonable.
\section{Appendix I}
\label{app}
In this appendix I provide the details of the calculation of the zeta function in the high temperature limit for the case of Dirichlet boundary conditions and timelike asymmetry, introduced in subsection \ref{3_2}.

The uniform density part of the zeta function, $\zeta_U(s)$, contains a piece that does not depend on the temperature
\begin{equation}
\zeta_{U,0}(s)={\mu^{2s}(1+\lambda)\over\Gamma(s)}{L^2a\over 8\pi^{3\over 2}}\int_0^\infty dt\, t^{s-{5\over 2}}{e^{-M^2t}F(eBt)},
\label{zetaU0}
\end{equation}
and a piece dependent on $T$
\begin{equation}
\zeta_{U,T}(s)={\mu^{2s}(1+\lambda)\over\Gamma(s)}{L^2a\over 4\pi^{3\over 2}}\int_0^\infty dt\, t^{s-{5\over 2}}{e^{-M^2t}F(eBt)}\sum_{m=1}^\infty  e^{-(1+\lambda){4\pi^2 m^2\over \beta^2}t}.
\label{zetaUT}
\end{equation}
After a change of integration variable from $t$ to $t/eB$, the integral in $\zeta_{U,0}(s)$ can be written as
\begin{equation}
\int_0^\infty dt\, t^{s-{5\over 2}}{e^{-M^2t}F(eBt)}=(eB)^{{3\over2}-s}\int_0^\infty dt\, t^{s-{3\over 2}}\sum_{l=0}^\infty e^{-(2l+1+{M^2\over eB})t},
\label{zetaU02}
\end{equation}
and, after another change of integration variable,
\begin{equation}
(eB)^{{3\over2}-s}\int_0^\infty dt\, t^{s-{3\over 2}}\sum_{l=0}^\infty e^{-(2l+1+{M^2\over eB})t}={\sqrt{2}(eB)^{3\over2}\over (2eB)^s}\zeta_H\left(s-{\textstyle\frac{1}{2}} ; {\textstyle\frac{1}{2}}+{\textstyle\frac{M^2}{2eB}}\right)\Gamma(s-{\textstyle\frac{1}{2}}),
\label{zetaU03}
\end{equation}
where $\zeta_H(s;z)$ is the Hurwitz zeta function defined as
\begin{equation}
\zeta_H(s;z) = \sum_{n=0}^\infty (n+z)^{-s}.
\label{Hurwitz}
\end{equation}
Making use of Eqs. (\ref{zetaU02}) and (\ref{zetaU03}), I obtain
\begin{equation}
\zeta_{U,0}(s)=\left({\mu^{2}\over 2eB}\right)^s{(1+\lambda)\over\Gamma(s)}{L^2a(eB)^{3\over2}\over 2^{5\over 2}\pi^{3\over 2}}\zeta_H\left(s-{\textstyle\frac{1}{2}} ; {\textstyle\frac{1}{2}}+{\textstyle\frac{M^2}{2eB}}\right)\Gamma(s-{\textstyle\frac{1}{2}}).
\label{zetaU04}
\end{equation}
Using the following power series expansion, valid for $s\ll 1$,
\begin{equation}
A^{s}{\Gamma(s-{\textstyle\frac{1}{2}})\over\Gamma(s)}\zeta_H\left(s-{\textstyle\frac{1}{2}} ; {\textstyle\frac{1}{2}}+{\textstyle\frac{z}{2}}\right)\simeq -2{\sqrt{\pi}}\zeta_H\left(-{\textstyle\frac{1}{2}} ; {\textstyle\frac{1}{2}}+{\textstyle\frac{z}{2}}\right)s +{\cal O}(s^2),
\label{app04}
\end{equation}
I easily evaluate $\zeta'_{U,0}(0)$ and include that contribution into $E_U$ of Eq. (\ref{EU_1}).

To evaluate the temperature dependent part $\zeta_{U,T}(s)$ I make the following approximations
\begin{equation}
e^{-M^2t}\simeq 1 -M^2t + {M^4t^2\over 2},
\label{app1}
\end{equation}
and 
\begin{equation}
F(eBt)\simeq 1 - {e^2B^2t^2\over 6},
\label{app2}
\end{equation}
valid since $M, \sqrt{eB} \ll  T$. Using these two approximations into Eq. (\ref{zetaU_1}), changing the integration variable from $t$ to $t\beta^2\over(1+\lambda)4\pi^2m^2$ and evaluating the integral, I obtain
\begin{eqnarray}
\zeta_{U,T}(s)&=&\!\!\!\
\left({\mu\beta\over\sqrt{1+\lambda}2\pi}\right)^{2s}{2\pi^{3\over 2}(1+\lambda)^{5\over 2}\over \Gamma (s)}T^3L^2a\left[\zeta_R(2s-3)\Gamma(s-{3\over 2})-{M^2\beta^2\over 4\pi^2(1+\lambda)}\zeta_R(2s-1)\Gamma(s-{1\over 2})\right.
\nonumber \\
&+&\!\!\!\
\left.\left({M^4\over 2}-{e^2B^2\over 6}\right){\beta^4\over 16\pi^4(1+\lambda)^2}\zeta_R(2s+1)\Gamma(s+{1\over 2}),
\right]
\label{app3}
\end{eqnarray}
where $\zeta_R(s)$ is the Riemann zeta function. Using the following power series expansions, valid for $s\ll 1$,
\begin{equation}
A^{2s}{\Gamma(s-{\textstyle\frac{3}{2}})\over\Gamma(s)}\zeta_R(2s-3)\simeq {\sqrt{\pi}\over 90}s +{\cal O}(s^2),
\label{app4}
\end{equation}
\begin{equation}
A^{2s}{\Gamma(s-{\textstyle\frac{1}{2}})\over\Gamma(s)}\zeta_R(2s-1)\simeq {\sqrt{\pi}\over 6}s +{\cal O}(s^2),
\label{app5}
\end{equation}
\begin{equation}
A^{2s}{\Gamma(s+{\textstyle\frac{1}{2}})\over\Gamma(s)}\zeta_R(2s+1)\simeq {\sqrt{\pi}\over 2}+{\sqrt{\pi}}\left[\gamma_E+\ln\left({A\over 2}\right)\right]s +{\cal O}(s^2),
\label{app6}
\end{equation}
I obtain  $\zeta'_{U,T}(0)$ and, in turn, $E_U=-T\zeta'_U(0)$ in the high temperature limit, shown in Eq. (\ref{EU_1}).

I also separate the other part of the zeta function, $\zeta_{T, a}(s)$, in two pieces. One piece 
\begin{equation}
\zeta_{a}(s)={\mu^{2s}(1+\lambda)\over\Gamma(s)}{L^2a\over 4\pi^{3\over 2}}\int_0^\infty dt\, t^{s-{5\over 2}}{e^{-M^2t}F(eBt)}\sum_{n=1}^\infty e^{-{n^2a^2\over t}}.
\label{app7}
\end{equation}
depends on $a$ and does not depend on $T$, the other one
\begin{equation}
\tilde{\zeta}_{T,a}(s)={\mu^{2s}(1+\lambda)\over\Gamma(s)}{L^2a\over 2\pi^{3\over 2}}\int_0^\infty dt\, t^{s-{5\over 2}}{e^{-M^2t}F(eBt)}\sum_{m=1}^\infty  e^{-(1+\lambda){4\pi^2 m^2\over \beta^2}t}\sum_{n=1}^\infty e^{-{n^2a^2\over t}}.
\label{app8}
\end{equation}
depends on both $a$ and $T$. I evaluate $\zeta_a(s)$ first. When $a^{-1} \gg M, \sqrt{eB}$, I use Eqs. (\ref{app1}) and (\ref{app2}), integrate, and find
\begin{eqnarray}
\zeta_{a}(s)&=&\!\!\! {(\mu a)^{2s}(1+\lambda)\over\Gamma(s)}{L^2\over 4\pi^{3\over 2}a^2}\left[\Gamma\left({\textstyle\frac{3}{2}}-s\right)\zeta_R(3-2s)-M^2a^2\Gamma\left({\textstyle\frac{1}{2}}-s\right)\zeta_R(1-2s)
\right.
\nonumber \\
&+&\!\!\!
\left.\left({M^4\over 2}-{e^2B^2\over 6}\right)a^4
\Gamma\left(-{\textstyle\frac{1}{2}}-s\right)\zeta_R(-1-2s)\right].
\label{app9}
\end{eqnarray}
Using the following expansions,
\begin{equation}
A^{2s}{\Gamma({\textstyle\frac{3}{2}}-s)\over\Gamma(s)}\zeta_R(3-2s)\simeq {\sqrt{\pi}\zeta_R(3)\over 2}s+{\cal O}(s^2),
\label{app10}
\end{equation}
\begin{equation}
A^{2s}{\Gamma({\textstyle\frac{1}{2}}-s)\over\Gamma(s)}\zeta_R(1-2s)\simeq -{\sqrt{\pi}\over 2}\left(s^{-1}+2\ln 2A\right)s+{\cal O}(s^2),
\label{app11}
\end{equation}
\begin{equation}
A^{2s}{\Gamma(-{\textstyle\frac{1}{2}}-s)\over\Gamma(s)}\zeta_R(-1-2s)\simeq {\sqrt{\pi}\over 6}s +{\cal O}(s^2),
\label{app12}
\end{equation}
where $\zeta_R(3)=1.2021$, I obtain $\zeta_{a}(s)$ in the small $s$ limit for $a^{-1} \gg M, \sqrt{eB}$
\begin{equation}
\zeta_{a}(s)={(1+\lambda)L^2\over 8\pi a^2}\left[\zeta_R(3)+M^2a^2\left(s^{-1}+2\ln 2\mu a\right)+\left({M^4\over 6}-{e^2B^2\over 18}\right)a^4
\right]s.
\label{app13}
\end{equation}
The contribution of $\zeta_{a}(s)$ to the Casimir energy, $E_a=-T\zeta'(0)$, is 
\begin{equation}
E_{a}=-{(1+\lambda)L^2T\over 8\pi a^2}\left[\zeta_R(3)+2M^2a^2\ln (2M a)+\left({M^4\over 6}-{e^2B^2\over 18}\right)a^4
\right],
\label{Ea1}
\end{equation}
where I made the obvious choice $\mu = M$. When $\sqrt{eB}\gg a^{-1}, M$, I write $F(eBt)$ in Eq. (\ref{app7}) as a sum of exponentials using Eq. (\ref{sinh}) and change integration variable from $t$ to ${tna\over\sqrt{M^2+(2\ell+1)eB}}$. Only the term with $n=1$ and $\ell=0$ contributes significantly to the double sum and, once I integrate using the saddle point method, I find
\begin{equation}
\zeta_{a}(s)={(1+\lambda)L^2eB\over 2\pi \Gamma(s)}\left({a\mu^2\over \sqrt{M^2+eB}}\right)^s e^{-2a\sqrt{M^2+eB}},
\label{app14}
\end{equation}
whose contribution to the Casimir energy is
\begin{equation}
E_{a}=-{(1+\lambda)L^2TeB\over 2\pi} e^{-2a\sqrt{M^2+eB}}.
\label{Ea2}
\end{equation}
When $M\gg a^{-1}, \sqrt{eB}$, I change variable of integration from $t$ to ${tna\over M}$ and only the term with $n=1$ contributes significantly. I integrate and find
\begin{equation}
\zeta_{a}(s)=\left({a\mu^2\over M}\right)^s {(1+\lambda)L^2M\over 4\pi a\Gamma(s)}e^{-2Ma}F\left({eBa\over M}\right),
\label{app14a}
\end{equation}
whose contribution to the Casimir energy is reported below
\begin{equation}
E_{a}=-{(1+\lambda)L^2TM\over 4\pi a} e^{-2Ma}F\left({eBa\over M}\right).
\label{Ea2a}
\end{equation}

Finally, I evaluate $\tilde{\zeta}_{T,a}(s)$ by changing the integration variable in Eq. (\ref{app8}) from $t$ to $tna\over 2\pi m\sqrt{(1+\lambda)}T$. Only the term with $n=m=1$ contributes significantly and I integrate it using the saddle point method, to obtain
\begin{equation}
\tilde{\zeta}_{T,a}(s)=\left(\mu^2 a\over 2\pi T\sqrt{1+\lambda}\right)^{s}{(1+\lambda)^{\frac{3}{2}}L^2\over\Gamma(s)}{T\over a}e^{-{4\pi  a T \sqrt{1+\lambda}}}e^{-{M^2a\over 2\pi T \sqrt{1+\lambda}}}F\left({eBa\over 2\pi T \sqrt{1+\lambda}}\right).
\label{app15}
\end{equation}
The contribution of $\tilde{\zeta}_{T,a}(s)$ to the Casimir energy is now quickly obtained and reported in Eq. (\ref{ETa1}). 
\section{Appendix II}
\label{appII}
In this appendix I show details of the calculation of the high temperature zeta function for mixed boundary conditions and timelike anysotropy, a topic I address in Section \ref{4}.
The uniform energy density piece of this zeta function is identical to that obtained in Eqs. (\ref{zetaU03}) and (\ref{app3}) of Appendix I for Dirichlet boundary conditions. The remaining part of the zeta function is again separated into a part that depends on $a$ and not on $T$, 
\begin{equation}
\zeta_{a}(s)={\mu^{2s}(1+\lambda)\over\Gamma(s)}{L^2a\over 4\pi^{3\over 2}}\int_0^\infty dt\, t^{s-{5\over 2}}{e^{-M^2t}F(eBt)}\sum_{n=1}^\infty (-1)^n e^{-{n^2a^2\over t}},
\label{app2_1}
\end{equation}
and a part that depends on both $a$ and $T$, 
\begin{equation}
\tilde{\zeta}_{T,a}(s)={\mu^{2s}(1+\lambda)\over\Gamma(s)}{L^2a\over 2\pi^{3\over 2}}\int_0^\infty dt\, t^{s-{5\over 2}}{e^{-M^2t}F(eBt)}\sum_{m=1}^\infty  e^{-(1+\lambda){4\pi^2 m^2\over \beta^2}t}\sum_{n=1}^\infty (-1)^ne^{-{n^2a^2\over t}}.
\label{app2_2}
\end{equation}
I evaluate $\zeta_a(s)$ first and, as I did in the previous appendix, consider three asymptotic cases, $a^{-1}\gg\sqrt{eB}, M$; $\sqrt{eB} \gg a^{-1}, M$; and $M\gg\sqrt{eB}, a^{-1}$. In the case of $a^{-1}\gg\sqrt{eB}, M$, I use Eqs. (\ref{app1}) and (\ref{app2}), integrate and obtain
\begin{eqnarray}
\zeta_{a}(s)&=&\!\!\! {(\mu a)^{2s}(1+\lambda)\over\Gamma(s)}{L^2\over 4\pi^{3\over 2}a^2}\left[\Gamma\left({\textstyle\frac{3}{2}}-s\right)(4^{s-1}-1)\zeta_R(3-2s)-M^2a^2\Gamma\left({\textstyle\frac{1}{2}}-s\right)(4^{s}-1)\zeta_R(1-2s)
\right.
\nonumber \\
&+&\!\!\!
\left.\left({M^4\over 2}-{e^2B^2\over 6}\right)a^4
\Gamma\left(-{\textstyle\frac{1}{2}}-s\right)(4^{s+1}-1)\zeta_R(-1-2s)\right],
\label{app2_3}
\end{eqnarray}
and, using the following power series expansions valid for $s\ll 1$
\begin{equation}
A^{2s}{\Gamma({\textstyle\frac{3}{2}}-s)\over\Gamma(s)}(4^{s-1}-1)\zeta_R(3-2s)\simeq -{3\sqrt{\pi}\over 8}\zeta_R(3)s+{\cal O}(s^2),
\label{app2_4}
\end{equation}
\begin{equation}
A^{2s}{\Gamma({\textstyle\frac{1}{2}}-s)\over\Gamma(s)}(4^{s}-1)\zeta_R(1-2s)\simeq -{\sqrt{\pi}}\ln (2) s+{\cal O}(s^2),
\label{app2_5}
\end{equation}
\begin{equation}
A^{2s}{\Gamma(-{\textstyle\frac{1}{2}}-s)\over\Gamma(s)}(4^{s+1}-1)\zeta_R(-1-2s)\simeq {\sqrt{\pi}\over 2}s +{\cal O}(s^2),
\label{app2_6}
\end{equation}
I write $\zeta_a(s)$ in the small $s$ limit as
\begin{equation}
\zeta_{a}(s)\simeq{(1+\lambda)}{L^2\over 4\pi a^2}\left[-{3\over 8}\zeta_R(3)+\ln(2)M^2a^2
+\left({M^4\over 4}-{e^2B^2\over 12}\right)a^4
\right]s+{\cal O}(s^2),
\label{app2_7}
\end{equation}
and immediately obtain its contribution $E_a=-T\zeta'_a(0)$ to the Casimir energy, which I report in Eq. (\ref{Ea7}) of Section \ref{4}.
When $\sqrt{eB} \gg a^{-1}, M$  I write $F(eBt)$ as a sum of exponentials and change integration variable from $t$ to ${tna\over\sqrt{M^2+(2\ell+1)eB}}$. Only the term with $n=1$ and $\ell=0$ contributes significantly to the double sum and, once I do the integration using the saddle point method, I find
\begin{equation}
\zeta_{a}(s)=-{(1+\lambda)L^2eB\over 2\pi \Gamma(s)}\left({a\mu^2\over \sqrt{M^2+eB}}\right)^s e^{-2a\sqrt{M^2+eB}},
\label{app2_8}
\end{equation}
whose contribution to the Casimir energy is reported  in Eq. (\ref{Ea8}) of Section \ref{4}. When $M\gg\sqrt{eB}, a^{-1}$ I change variable of integration from $t$ to ${tna\over M}$, retain the only significant term with $n=1$, integrate and find
\begin{equation}
\zeta_{a}(s)=-{(1+\lambda)L^2M\over 4\pi a\Gamma(s)}\left({a\mu^2\over M}\right)^s e^{-2Ma}F\left({eBa\over M}\right),
\label{app2_8a}
\end{equation}
whose contribution to the Casimir energy is reported  in Eq. (\ref{Ea8a}) of Section \ref{4}.

Finally, $\tilde\zeta_{T,a}$ is computed with a change of integration variable, as described in Appendix I.
The result is similar to Eq. (\ref{app15}) of Appendix I, the only difference being an overall negative sign.

\end{document}